\documentclass[aps,prl,preprint,groupedaddress]{revtex4}
\usepackage{amssymb}
\usepackage{graphicx}

\begin{document}

\title{Anomalous negative differential thermal resistance in a
momentum-conserving lattice}
\author{Wei-Rong Zhong $^{1}$}
\email{wrzhong@hotmail.com}
\author{Mao-Ping Zhang $^{1}$}
\author{Bao-Quan Ai $^{2}$}
\email{aibq@scnu.edu.cn}
\author{Bambi Hu $^{3}$}
\affiliation{$^{1}$\textit{Department of Physics and Siyuan Laboratory, College of
Science and Engineering, Jinan University, Guangzhou, 510632, China.}}
\affiliation{$^{2}$\textit{Laboratory of Quantum Information Technology, ICMP and SPTE,
South China Normal University, Guangzhou, 510006 P. R. China.}}
\affiliation{$^{3}$\textit{Department of Physics, University of Houston, Houston, Texas
77204-5005, USA.}}
\date{\today }

\begin{abstract}
A two-segment Fermi-Pasta-Ulam lattices has been investigated by using
nonequilibrium molecular dynamics. Here we present an anomalous negative
differential thermal resistance (NDTR) that have not been reported in
Frenkel-Kontorova and $\phi _{4}$ lattices up to the present. The NDTR
disappears at low temperature region. The region of NDTR shifts from the
large to the small temperature difference region as the system size
increases. Anomalous dependence of NDTR on the temperature can be explained
as the negative effect induced by the nonlinear coupling. The explanation
can also cover the phenomenon of NDTR in momentum-nonconserved lattices.
\end{abstract}

\maketitle

\textbf{I. INTRODUCTIONS}

Negative differential thermal resistance (NDTR), which may implies an
intrinsic physical mechanism for some thermal devices, has attracted lots of
attentions in recent years. NDTR is first found in asymmetric
Frenkel-Kontorova (FK) lattices \cite{bli}, and then in pure anharmonic
systems \cite{he}. Analytical and numerical studies have been carried to
reveal the original property of NDTR \cite%
{bli,he,ai,zhong1,shao,segal,pereira}. Up to now NDTR has been found in many
low-dimensional systems, such as the anharmonic lattice with gradient mass
\cite{nyang}, two FK lattice with weak link \cite{bli, zhong1}, the pure $%
\phi _{4}$ and FK lattice \cite{he,ai}, the double-stranded systems \cite%
{zhong2}, and so on. It has been reported that NDTR depends on the nonlinear
external potential, the finite size and the temperature of the system.

The main explanations about the phenomenon of NDTR include several points:
1) the mismatch of the phonon spectra of the two particles \cite{bli,nyang};
2) the competition between the temperature difference, which acts as an
external field, and the temperature-dependent thermal boundary conductance
\cite{he}; 3) the effective phonon-band shifts \cite{shao}; 4) a ballistic
transport that induce the competition between the molecular occupation
factor and the temperature difference \cite{zhong1}; 5) the phonon-lattice
scattering becomes so significant that NDTR occurs \cite{he,ai}. All these
explanations suggest a common view that the NDTR cannot occur when the
system size increases, or in the system without nonlinear external
potential, or in the system as the temperature is increasing. In other
words, NDTR only appears at low temperature in a small system with nonlinear
external potential. For the sake of comparison, here we call this kind of
NDTR the normal NDTR. According to the normal NDTR, it is proposed that the
momentum-conserved system without nonlinear external potential, such as a
Fermi-Pasta-Ulam (FPU) lattice, behaves no NDTR phenomenon. Since the
nano-materials such as the carbon nanotubes is a momentum-conserved system
\cite{bli2, tabar}, this view about no NDTR in the momentum-conserved system
once places the studies of the NDTR in carbon nanotubes in a difficult
position. Actually, it is not the case.

In this paper, a two-segment FPU lattice with a weak link is studied by
molecular dynamics simulations. We will present that the NDTR can also occur
in a momentum-conserved system. It is surprising that the NDTR in FPU
lattice behaves an anomalous dependence on the temperature and the system
size, which is different from the NDTR found in the system with nonlinear
external potential. Here we introduce a competition between the linearity
and the nonlinearity to interpret the origin of NDTR. Our explanation can
cover some of the phenomena of NDTR that have been reported in FK models.

\textbf{II. MOMENTUM-CONSERVING AND MOMENTUM-NONCONSERVING LATTICES}

\textit{Momentum-conserving lattice} \cite{lepri} A chain of N coupled
atoms, in which only nearest-neighbor interactions will be considered for
simplicity. The first class of models we wish to consider are defined by an
Hamiltonian of the form ($p_{l}=m_{l}x_{l}$)

\begin{equation}
\varkappa =\sum_{l=1}^{N_{M}}\left[ \frac{p_{l}^{2}}{2m_{l}}+V(x_{l+1}-x_{l})%
\right] ,
\end{equation}

Boundary conditions need also to be specified by defining $x_{0}$ and $%
x_{N+1}$. Typical choices are periodic, fixed or free boundaries. As only
internal forces, which depend on relative positions, are present, the total
momentum is conserved and thus a zero mode exist. The important examples are
the well-known Lennard--Jones potential and Fermi--Pasta--Ulam (FPU)
potential.

\textit{Momentum-nonconserving lattice} \cite{lepri} At the simplest level
of modelization, this can be described by adding an external, on-site
potential to Eq.(1). For instance, neglecting the transverse motion leads to
one-dimensional models of the form:

\begin{equation}
\varkappa =\sum_{l=1}^{N_{M}}\left[ \frac{p_{l}^{2}}{2m_{l}}%
+U(x_{l})+V(x_{l+1}-x_{l})\right] ,
\end{equation}

The substrate potential $U(x_{l})$ breaks the invariance $x_{l}\rightarrow
x_{l}+const$: of Eq.(2) and the total momentum is no longer a constant of
the motion. An important example is the well-known Frenkel-Kontorova (FK)
potential.

\textbf{III. MODELS AND SIMULATIONS}

The nonlinear lattices that we use in this letter consist of two segments,
left segment (L) and right segment (R). Each segment is a FPU lattice.
Segment L and R are coupled via a spring of constant $k_{int}$. The total
Hamiltonian of the model is%
\begin{equation}
H=H_{L}+H_{R}+H_{int},
\end{equation}%
and the Hamiltonian of each segment can be written as%
\begin{equation}
H_{M}=\sum_{i=1}^{N_{M}}\left[ \frac{p_{M,i}^{2}}{2m_{M}}+\frac{k_{M}}{2}%
\left( x_{M,i+1}-x_{M,i}\right) ^{2}+\frac{\beta _{M}}{4}\left(
x_{M,i+1}-x_{M,i}\right) ^{4}\right] ,
\end{equation}%
with $x_{M,i}$ and $x_{M,i}$ denote the displacement from equilibrium
position and the conjugate momentum of the $i^{th}$ particle in segment $M$,
where $M$ stands for $L$ or $R$. The parameters $k_{M}$ and $\beta _{M}$ are
the harmonic and anharmonic spring constant of the FPU lattice,
respectively. We couple the last particle of segment $L$ and $R$ via a
harmonic spring. Thus, $H_{int}=\frac{k_{int}}{2}\left(
x_{L,N}-x_{R,N}\right) ^{2}.$ For the sake of simplicity, we set the mass of
particle $m=1$ and the Boltzmann constant $k_{B}=1$.

In our simulations we use fixed boundary condition and the chain is
connected to two heat baths at temperature $T_{L}$ and $T_{R}$. It is
reported that the Nos\'{e}-Hoover thermostat may induce an inauthentic NDTR
in the system without nonlinear external potential \cite{nose, ai}. To avoid
the adverse effect of the Nos\'{e}-Hoover thermostat we use the Langevin
heat baths and integrate the equations of motion by using the Verlet
frog-jumping algorithm \cite{art, press}. The local temperature is defined
as $T_{i}=\left\langle p_{i}^{2}\right\rangle $. The local heat flux is
defined as $j_{i}=k_{M}\langle p_{i}(x_{i}-x_{i-1})\rangle +\beta
_{M}\langle p_{i}(x_{i}-x_{i-1})^{3}\rangle $, and the total heat flux is $%
J=Nj$. The simulations are performed long enough to allow the system to
reach a steady state in which the local heat flux is constant along the
chain. For the sake of comparison, we define a heat current ratio, $%
J_{R}=J/J_{\max }$, in which $J_{\max }$ is the maximum heat current under a
fixed temperature $T_{R}$ of the right heat bath. The transport coefficient
is an important quantity for characterizing the transport mode of a thermal
transport process \cite{jswang, dli}. The thermal conductance evaluated as $%
\sigma =Nj/\triangle T$ represents an effective transport coefficient that
includes both boundary and bulk resistances \cite{lepri}. Here we fix the
temperature of the hot heat baths and change the temperature of the cold
heat baths. The temperature difference increases with the decrease of the
temperature of cold heat baths.

\textbf{IV. RESULTS AND\ DISCUSSIONS}

Figure 1 shows the temperature dependence of NDTR in the two-segment FPU
lattice. In the lattices with nonlinear external potential, which also
called momentum-nonconserved systems, lots of studies have shown that the
NDTR disappears with the temperature increasing of the systems \cite{zhong1,
ai, shao, he}. However, here it is presented that the NDTR occurs with the
temperature increasing of the hot heat baths. Interestingly, the results for
momentum-conserved systems are contrary to those for momentum-nonconserved
systems.
\begin{figure}[htbp]
\begin{center}\includegraphics[width=8cm,height=6cm]{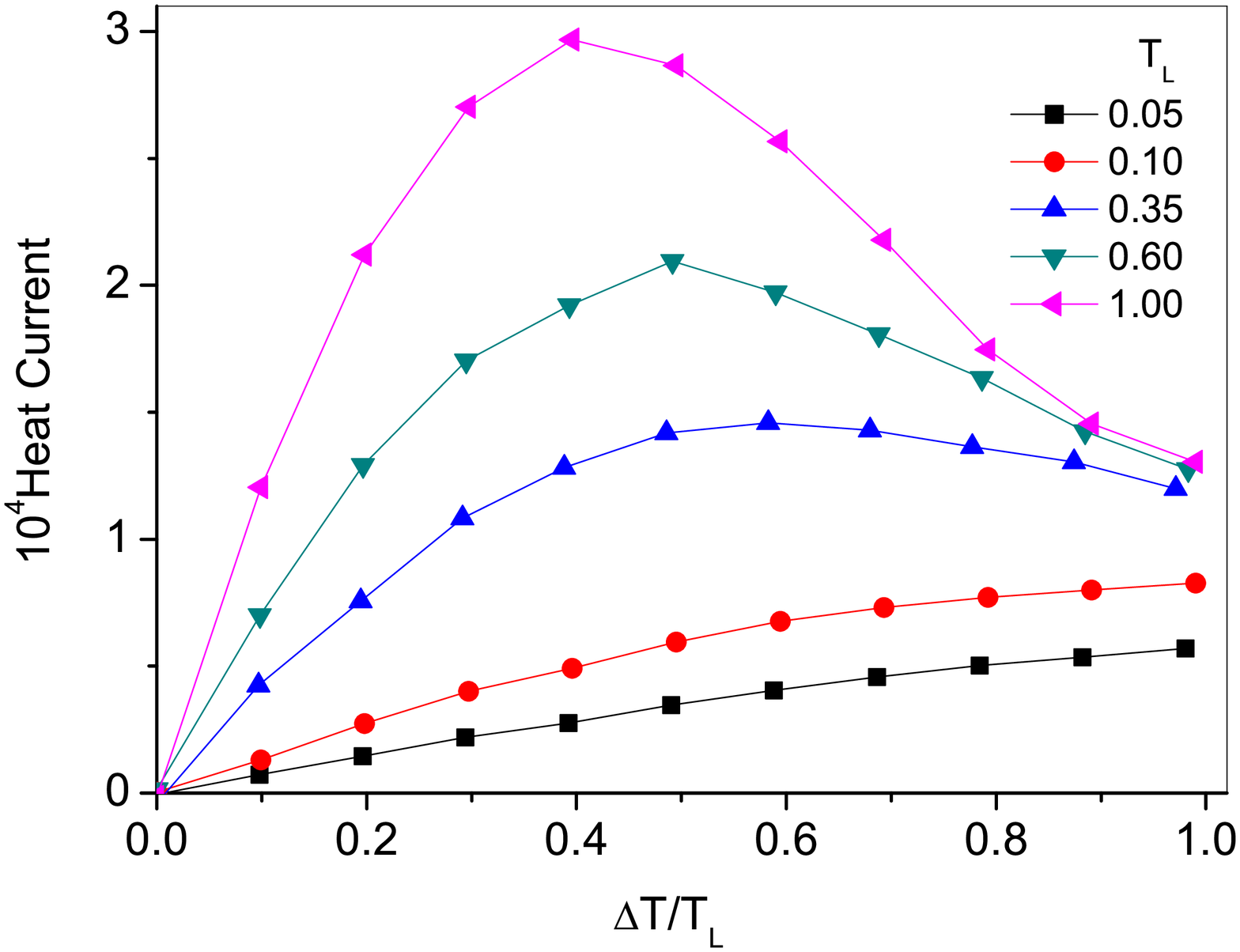}
  \end{center}
  \caption{Heat current as a
function of the temperature difference for various fixed high temperature $%
T_{L}$=0.05, 0.1, 0.35, 0.60, 1.0. The linear and nonlinear couplings are
respectively $k=1.0$ and $\protect\beta =0.5$, the weak link coupling $%
k_{int}$=0.01, and the system size is 64.}
   \label{}
\end{figure}

This anomalous NDTR can be understood from the competition between linear
and nonlinear interactions. Firstly we would like to discuss the role of
linear and nonlinear interaction on the thermal conductivity. It is well
known that positive effects of the linear interaction on the thermal
conductivity of the system have been confirmed by many studies, which
include not only numerical but also analytical results \cite{lepri, pereira,
bonetto}. It can be described as a function $K$\symbol{126}$\lambda ^{\alpha
}$, where $K$ and $\lambda $ are respectively thermal conductivity and the
linear coupling, the exponent $\alpha >0$. Correspondingly, it is also shown
that the nonlinear external potential preforms a negative influence on the
thermal conductivity \cite{lepri, pereira, bonetto}. The relationship is
written as $K$\symbol{126}$V^{\beta }$, in which $V$ is the strength of the
nonlinear coupling or the external potential, the exponent $\beta <0$. The
NDTR depends exactly on the competition between these negative and the
positive effects. When the negative effects surpass the positive ones, the
NDTR occurs. Otherwise, the DNTR disappears.

For the FPU lattices, figure2 shows that the heat current decreases with the
nonlinear coupling constant, which implies a nonlinear inhibition on the
thermal conductivity. In the absence of nonlinear coupling, i.e., in the
case of a harmonic lattice, as shown in Fig. 2, when the nonlinear coupling $%
\beta $ equals zero, the heat current increases linearly with the
temperature difference and no NDTR occurs. When $\beta $ increases to a
value larger than 0.3, due to the enhancement of the nonlinear effect the
NDTR appears at the region of the large temperature difference.
\begin{figure}[htbp]
\begin{center}\includegraphics[width=8cm,height=6cm]{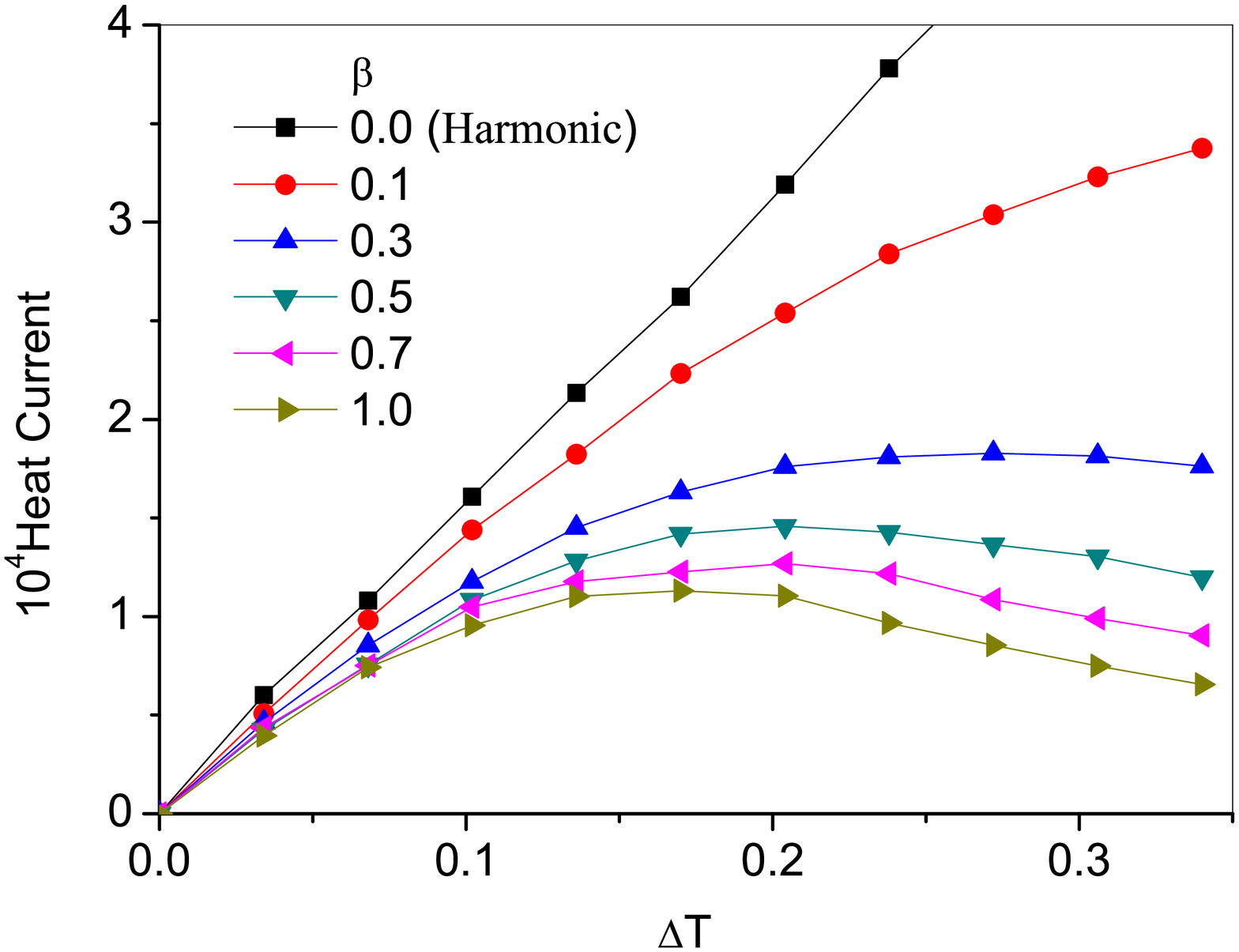}
  \end{center}
  \caption{Heat current as a function of temperature
difference for various nonlinear couplings. The temperature $T_{L}=0.35$,
and the remain parameters are the same as those for Fig.1.}
   \label{}
\end{figure}

In order to give more details for this explanation, we plot figure3 to show
the dependence of the linear and nonlinear forces on the displacement. As
mentioned above, in FPU lattice the NDTR disappears for low temperature and
occurs for high temperature. For the system with low temperature, the
particles oscillates near the equilibrium position and the displacement
between two nearest neighbor particle is small, i.e., $\Delta x_{i}\symbol{%
126}0$. As shown in Fig. 3(a), w
hen $\Delta x_{i}\symbol{126}0$, the force
from linear coupling ($-k\Delta x_{i}$) is larger than that from nonlinear
coupling ($-\beta \Delta x_{i}^{3}$). Therefore, the linear coupling
determinate the thermal transport and no NDTR occurs. As the temperature of
the system increases, $\Delta x_{i}|$ goes to a value
larger than $\Delta x_{C}|$ and the force from $-\beta
\Delta x_{i}^{3}$ is larger than $-k\Delta x_{i}$. The nonlinear coupling as
the negative effect determinate the thermal transport and then NDTR occurs.
However, in the FK lattices an opposite temperature dependence of NDTR is
observed. The Hamiltonian of FK model is $H_{FK}=\sum_{i=1}^{N}\left[ \frac{%
p_{i}^{2}}{2m}+\frac{k_{0}}{2}\left( x_{M,i+1}-x_{M,i}\right) ^{2}+\frac{V}{%
2\pi }(1-\cos (\frac{2\pi }{a}x_{i}))\right] ,$ where $x_{i}$ and $p_{i}$
denote the displacement from equilibrium position and the 
conjugate momentum
of the $i^{th}$ particle. The parameters $k_{0}$ and $V$ are the harmonic
spring constant and the strength of the external potential, respectively.
Here we can also give a explanation from the relationship between the force
and the displacement. As shown in Fig. 3(b), for the system with low
temperature, the force from linear coupling ($-k_{0}\Delta x_{i}$, $%
k_{0}=0.8 $) is smaller than that from nonlinear coupling ($-V\sin x_{i}$, $%
V=0.6$) and the nonlinear coupling determinate the thermal transport and
then NDTR occurs. On the contrary, for the FK lattices with high
temperature, the superior linear coupling induces the disappearance of the
NDTR. If one set $k_{0}=3.0$, the linear force is always larger than the
nonlinear force, then the NDTR disappears, which have been confirmed in Refs.%
\cite{he, ai}.

\begin{figure}[htbp]
\begin{center}\includegraphics[width=8cm,height=6cm]{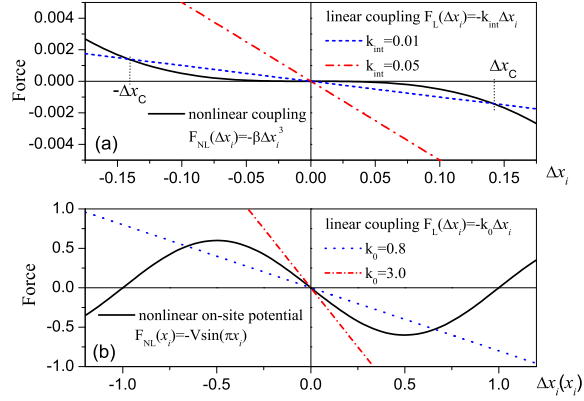}
  \end{center}
  \caption{Force depends on
the displacement of the particle. (a) linear and nonlinear coupling force in
FPU model, (b) linear coupling and nonlinear on-site force in FK model.}
   \label{}
\end{figure}

Figure 4 shows how the coupling constant $k_{int}$ affects the temperature
dependence of the heat flux $J/J_{\max }.$ The region of NDTR diminishes for
increasing $k_{int}$ until it vanishes at a critical coupling constant. The
weak link $k_{int}$ is a linear coupling, as also depicted in Fig.3(a), when
$k_{int}$ is small, the nonlinear effect can easily get above the linear one
at $\Delta x_{i}\symbol{126}0$ in the interface of the two FPU lattices and
then NDTR occurs. Instead, the NDTR vanishes for strong linear link $k_{int}$%
.

\begin{figure}[htbp]
\begin{center}\includegraphics[width=8cm,height=6cm]{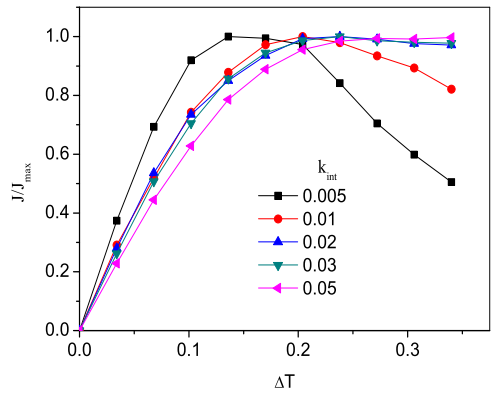}
  \end{center}
  \caption{Heat current as a function of
temperature difference for different weak link couplings $k_{int}$. Here the
remain parameters are k=1.0, $\protect\beta $=0.5, T$_{L}$=0.35 and N=64,
respectively.}
   \label{}
\end{figure}

In addition, the system size dependence of NDTR in FPU lattices is also
different from that have been found in FK and $\phi _{4}$ lattices. In FK
and $\phi _{4}$ lattices \cite{he, ai}, with the increasing of the system
size, the NDTR region shifts from the small to the large temperature
difference region and vanishes in the end. On the contrary, as shown in Fig.
5, in FPU lattices the NDTR disappears for the small system. When the system
size increases, the region of NDTR (the regime with a frame) shifts from the
large to the small temperature difference region. 

\begin{figure}[htbp]
\begin{center}\includegraphics[width=8cm,height=6cm]{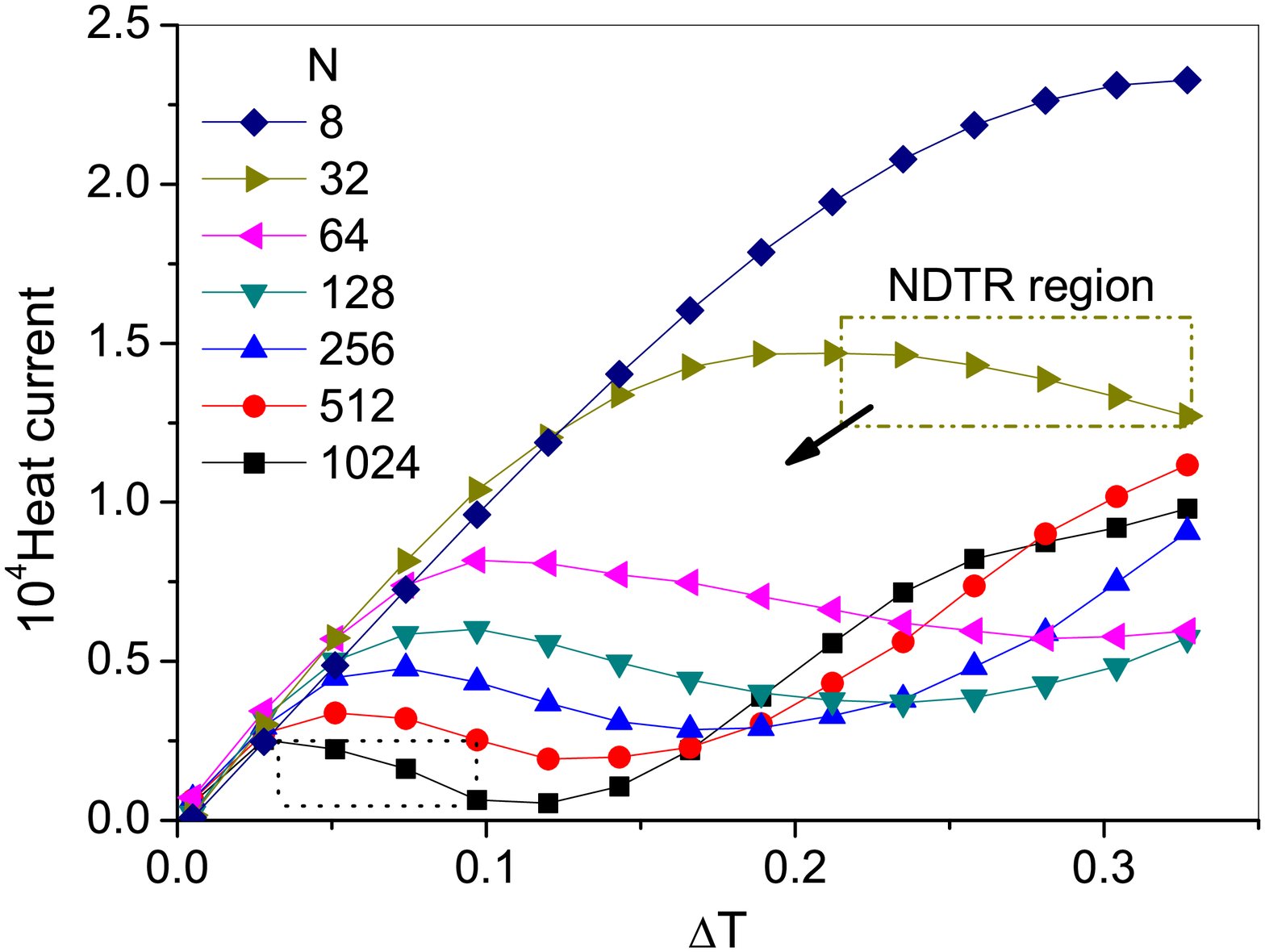}
  \end{center}
  \caption{NDTR depends on the system size. Here the remain
parameters are k=1.0, $\protect\beta $=0.5, T$_{L}$=0.35 and $k_{int}$=0.05,
respectively.}
   \label{}
\end{figure}

\textbf{V. COMPARISONS}

Here we would like to compare our explanations about NDTR with those have
been presented. As mentioned above, the NDTR have been explained as the
phonon-lattice scattering, the crossover of the transport mode, the
competition between the temperature difference and the boundary conductance,
the mismatch of phonon spectra and so on. As have been reported, the
mismatch of phonon spectra and the effective phonon-band shifts can give a
good interpretation for the NDTR in two-segment lattices, but fail to
interpret the NDTR in pure single lattice \cite{shao, he, ai}. Moreover, the
competition between the temperature difference and the temperature-dependent
thermal boundary conductance can only explain the small systems with strong
boundary effect \cite{he, ai}, but will be invalid for large systems
mentioned here. Finally, the phonon-lattice scattering and the crossover of
the transport mode approach the main reason of NDTR at low temperature \cite%
{he, ai, zhong1}, but not an appropriate origin of NDTR at high temperature.
Actually, all these explanations are based upon a negative effect, which
maybe imply a fundamental physical mechanism of NDTR. And here this negative
effect comes from the nonlinearity that includes both the nonlinear coupling
and the nonlinear external potential.

A comparison of NDTR in the two cases is also presented from three aspects:
1) \textit{the temperature dependence of NDTR}. In Ref. \cite{zhong1}, a
temperature dependence of NDTR is reported. When the temperature of hot heat
baths $T_{R}$ is small, there exists NDTR. As $T_{R}$ increases, the NDTR
disappears. However, here it is presented that the NDTR occurs with the
temperature increasing of the hot heat baths; 2) \textit{the system size
dependence of NDTR}. In the previous works (FK and phi\_4 model), as shown
in Refs. \cite{he, ai, shao}, NDTR becomes weak and disappears with the
increasing of the system size. The most important property is the NDTR
region shifts from the small to the large temperature difference region and
vanishes in the end. In the current works, on the contrary, as shown in Fig.
5, in FPU lattices the NDTR disappears for the small system. When the system
size increases, the region of NDTR (the regime with a frame) shifts from the
large to the small temperature difference region. 3) \textit{the nonlinear
potential}. As shown in Eqs. (1) and (2), the Hamiltonian of
momentum-conserving lattice is much different from that of
momentum-nonconserving lattice. Many reports have shown that the
momentum-conserving lattice has an anomalous thermal conductivity \cite%
{baowen}. Then it is naturally to regard the NDTR of momentum-conserving
lattice as an `anomalous' one. Additionally, the NDTR in this paper performs
very different or even opposite behavior from the NDTR that have been
reported in the previous papers, then we have to call it `anomalous NDTR'.

\textbf{VI. CONCLUSIONS AND APPLICATIONS}

In summary, we have reported an anomalous negative differential thermal
resistance in two-segment FPU chains through molecular dynamics simulations.
The NDTR occurs in the system when the system size as well as the
temperature increases. This anomalous NDTR effect can be understood from the
negative effect induced by the nonlinear force, which can give a more
fundamental explanation about NDTR. Since FPU lattice is a
momentum-conserved system, our results provide a new view about the NDTR in
the system without external potential, which is regarded as a necessary
condition for NDTR up to the present. Furthermore, our results have also
suggested that the NDTR can be achieved in two-segment nanoscale materials
with a weak linear link, which maybe a exciting information for fabricating
a nanoscale thermal device.

The experimental motivation of the model refers to the carbon nanotubes and
their extended models. The Lennard--Jones potential or Brenner--Tersoff
potential, which describes momentum-conserving lattice, is usually used to
study carbon nanotubes \cite{bli2, tabar}. A advanced structure of the
nanotubes is more similar to our model. For example, one can connect two
carbon nanotubes with silicon nanotubes. Here silicon nanotubes are weak
connections and carbon nanotubes are momentum-conserving lattices.

%\begin{figure}[htbp]
%\begin{center}\includegraphics[width=8cm,height=6cm]{fig1.eps}
% \end{center}
%\caption{}
% \label{}
%\end{figure}

\textbf{ACKNOWLEDGMENTS}

\begin{acknowledgments}
We would like to thank Siyuan clusters for running part of our programs.
This work was supported in part by the National Natural Science Foundation
of China (Grant Nos.11004082 and 11175067), the National Natural Science
Foundation of Guangdong Province, China (Grant Nos.10451063201005249 and
S2011010003323) and the Fundamental Research Funds for the Central
Universities, JNU (Grant No. 21609305). \
\end{acknowledgments}


\begin{thebibliography}{99}
\bibitem{bli} B. Li et al., Appl. Phys. Lett. \textbf{88}, 143501 (2006); L.
Wang and B. Li, Phys. Rev. Lett. \textbf{99}, 177208 (2007).

\bibitem{he} D. He, S. Buyukdagli, and B. Hu, Phys. Rev. B \textbf{80},
104302 (2009); D. H. He, B. Q. Ai, H. K. Chan and B. Hu, Phys. Rev. E
\textbf{81}, 041131 (2010).

\bibitem{segal} D. Segal and A. Nitzan, Phys. Rev. Lett. \textbf{94}, 034301
(2005); L.-A. Wu and D. Segal, ibid. \textbf{102}, 095503 (2009); D. Segal,
Phys. Rev. E \textbf{79}, 012103 (2009).

\bibitem{pereira} E. Pereira, Phys. Rev. E \textbf{82}, 040101(R) (2010); E.
Pereira and H. C. F. Lemos, ibid. \textbf{78}, 031108 (2008).

\bibitem{zhong1} W. R. Zhong, P. Yang, B. Q. Ai, Z. G. Shao and B. Hu, Phys.
Rev. E \textbf{79}, 050103(R) (2009).

\bibitem{ai} B. Q. Ai and B. Hu, Phys. Rev. E \textbf{83}, 011131 (2011); B.
Q. Ai, W. R. Zhong and B. Hu, Phys. Rev. E \textbf{83}, 052102 (2011).

\bibitem{shao} Z. G. Shao, L. Yang, H. K. Chan and B. Hu, Phys. Rev. E
\textbf{79}, 061119 (2009).

\bibitem{nyang} N. Yang, N. Li, L. Wang, and B. Li, Phys. Rev. B \textbf{76,}
020301(R) (2007).

\bibitem{zhong2} W. R. Zhong, Phys. Rev. E \textbf{81}, 061131 (2010).

\bibitem{bli2} G. Wu and B. Li, Phys. Rev. B 76, 085424 (2007).

\bibitem{tabar} H. Rafii-Tabar, Computational Physics of Carbon Nanotubes,
Cambridge University Press, New York(2008).

\bibitem{lepri} S. Lepri, R. Livi, and A. Politi, Phys. Rep. \textbf{377}, 1
(2003).

\bibitem{nose} S. Nose, J. Chem. Phys. \textbf{81}, 511 (1984); W. G.
Hoover, Phys. Rev.A \textbf{31}, 1695 (1985).

\bibitem{press} W. H. Press, S. A. Teukolsky, W. T. Vetterling, and B. P.
Flannery, Numerical Recipes (Cambridge University Press, Cambridge, 1992).

\bibitem{art} D. C. Rapaport, The Art of Molecular Dynamics Simulation,
Cambridge University Press, New York (2004).

\bibitem{dli} D. Li, Y. Wu, P. Kim, L. Shi, P. Yang, and A. Majumdara, Appl.
Phys. Lett. \textbf{83}, 2934 (2003).

\bibitem{jswang} J. S. Wang, Phys. Rev. Lett. \textbf{99}, 160601 (2007). Y.
Xu, J. S. Wang, W. Duan, B. L. Gu, and B. Li, Phys. Rev. B \textbf{78},
224303 (2008).

\bibitem{bonetto} F. Bonetto, J. L. Lebowitz, J. Lukkarinen, and S. Olla, J.
Stat. Phys. \textbf{134}, 1097 (2009); F. Bonetto, J. L. Lebowitz, and J.
Lukkarinen, J. Stat. Phys. \textbf{116}, 783 (2004); D. Roy, Phys. Rev. E
\textbf{77}, 062102 (2008).

\bibitem{baowen} O. Narayan and S. Ramaswamy, Phys. Rev. Lett, 89, 200601
(2002); B. Li and J. Wang, Phys. Rev. Lett, 91, 044301(2003).
\end{thebibliography}
\end{document}